\documentclass[12pt]{article}
\usepackage{epsfig,latexsym,amssymb,amsmath}
\begin{document}

\begin{center}
{\large \bf Stability analysis and quasinormal modes of Reissner Nordstr{\o}m Space-time
via Lyapunov exponent  }
\end{center}

\vskip 5mm

\begin{center}
{\Large{Parthapratim Pradhan\footnote{E-mail: pppradhan77@gmail.com}}}
\end{center}

\vskip  0.5 cm

{\centerline{\it Department of Physics}}
{\centerline{\it Vivekananda Satavarshiki Mahavidyalaya}}
%{\centerline{\it (Affiliated to Vidyasagar University)}}
{\centerline{\it Manikpara, Jhargram}}
{\centerline{\it West Midnapur~721513, India}}

\vskip 1cm

\begin{abstract}
We explicitly derive the proper time $(\tau)$ principal Lyapunov exponent ($\lambda_{p}$)
and coordinate time ($t$) principal Lyapunov exponent ($\lambda_{c}$)  for Reissner
Nordstr{\o}m (RN) black hole (BH) . We also compute their ratio.  For RN space-time, it is shown that
the ratio is  $\frac{\lambda_{p}}{\lambda_{c}}=\frac{r_{0}}{\sqrt{r_{0}^2-3Mr_{0}+2Q^2}}$ for
time-like circular geodesics  and for Schwarzschild BH it is
$\frac{\lambda_{p}}{\lambda_{c}}=\frac{\sqrt{r_{0}}}{\sqrt{r_{0}-3M}}$.
We  further show that their ratio $\frac{\lambda_{p}}{\lambda_{c}}$
may vary from orbit to orbit.
For instance,  Schwarzschild BH at innermost stable circular orbit(ISCO), the ratio is
$\frac{\lambda_{p}}{\lambda_{c}}\mid_{r_{ISCO}=6M}=\sqrt{2}$ and at
marginally bound circular orbit (MBCO) the ratio is calculated
to be $\frac{\lambda_{p}}{\lambda_{c}}\mid_{r_{mb}=4M}=2$.
Similarly, for  extremal RN BH the ratio at ISCO is
$\frac{\lambda_{p}}{\lambda_{c}}\mid_{r_{ISCO}=4M}=\frac{2\sqrt{2}}{\sqrt{3}}$.
We also  further analyse  the geodesic stability via this exponent. By evaluating the
Lyapunov exponent, it is shown that  in the eikonal limit , the real and
imaginary parts of the quasi-normal modes of RN BH is given by the frequency
and instability time scale of the unstable null circular geodesics.
\end{abstract}

%\keywords{ISCO, Lyapunov exponent, proper time, coordinate time, QNM.}
%\pacs{04.70.-s }

\section{Introduction}

Nonlinearity of Einstein's equation is the reason for non-linearity of Einstein's general theory of relativity. 
So, there may be a certain link between nonlinear Einstein's general theory of relativity and non-linear dynamics. 
Particularly, Lyapunov exponent \cite{lya} is one of the bridges between them. In this paper, we shall focus on 
analytical calculations involving the Lyapunov exponent in terms of the equation of circular geodesics around a 
BH space-time. These equatorial circular geodesics around a BH space-time playing a crucial role in general 
relativity for classification of the orbits. They also determine important features of the space-time and 
give the important information on the back ground geometry.

The Lyapunov exponent ($\lambda$) has been used to probe the instability of circular null geodesics and in 
terms of the quasi-normal modes (QNMs) for  spherically symmetric space-time of arbitrary dimensions \cite{car}, 
but the focus there is on null circular geodesics. It has been shown in ref. \cite{car} that in the eikonal 
approximation, the real and imaginary parts of the QNMs of any dimensions of  spherically 
symmetric, asymptotically flat space-time are given by (multiples of) the frequency and instability time 
scale of the unstable circular photon geodesics.

Note however that the principal Lyapunov exponents have been computed in \cite{car,clv,cl} using a 
\emph{co-ordinate} time $t$,  where $t$ is measured by the asymptotic observers. Thus, these exponents 
are explicitly coordinate dependent and therefore have a degree of un-physicality. Here, we compute the 
principal Lyapunov exponent analytically by using the \emph{proper time} as well as coordinate time and 
prove that their ratio i.e. $\frac{\lambda_{p}}{\lambda_{c}}$  is not an invariant quantity. Then we 
compare the results obtained both using the coordinate time and the proper time.  Using $\lambda$,  we 
also study the stability properties  of the  equatorial circular geodesics for RN BH space-time.

We  further elucidate the connection between  the Lyapunov exponent of the circular null geodesics  
in terms of the frequency of QNMs for RN BH in the eikonal limit. 
Interestingly, for \emph{extremal BH}  this frequency goes to zero. i.e.,
\begin{eqnarray}
\omega_{QNM}=0 ~.\label{nqn}
\end{eqnarray}
and for \emph{non-extremal RN BH} the frequency of QNM is
\begin{eqnarray}
\omega_{QNM} &=& \ell \sqrt{\frac{Mr_{c}-Q^{2}}{r_{c}^{4}}} - i\left(n+1/2\right)
\sqrt{\frac{(Mr_{c}-Q^{2})(3Mr_{c}-4Q^{2})}{r_{c}^{6}}}   ~.\label{nqn1}
\end{eqnarray}
where, $n$ is the overtone number, $\ell$ is the angular momentum of the perturbation.

For Schwarzschild BH, the  QNM frequency becomes
\begin{eqnarray}
\omega_{QNM} &=& \ell \sqrt{\frac{M}{r_{c}^{3}}} - i\left(n+1/2\right) \frac{\sqrt{3}M}{r_{c}^{2}} ~.\label{nqs}
\end{eqnarray}

Another interesting feature we have studied here is that,  when we use the proper time, the principal Lyapunov
exponent for RN space-time  can be obtained as
\begin{eqnarray}
\lambda_{p} &=& \sqrt{\frac{-(Mr_{0}^{3}-6M^{2}r_{0}^{2}+9MQ^{2}r_{0}-4Q^{4})}
{r_{0}^{4}(r_{0}^{2}-3Mr_{0}+2Q^{2})}} ~.\label{rnp}
\end{eqnarray}

When we use the coordinate time, the Lyapunov exponent for RN space-time
can be obtained
\begin{eqnarray}
\lambda_{c} &=& \sqrt{\frac{-(Mr_{0}^{3}-6M^{2}r_{0}^{2}+9M Q^{2}r_{0}-4Q^{4})}
{r_{0}^{6}}}~.\label{phr}
\end{eqnarray}
and  also their ratio is
\begin{eqnarray}
\frac{\lambda_{p}}{\lambda_{c}} &=& \frac{r_{0}}{\sqrt{r_{0}^2-3M r_{0}+2Q^2}} ~.\label{rnpc}
\end{eqnarray}
not an invariant quantity .

We would like to mention here for the reader a few works that have addressed
Lyapunov exponent  for different types of BHs. The invariant properties of Lyapunov exponent
were first discussed in \cite{karas} (see also \cite{mott,xu}). Sota et al. \cite{sota},
have first proposed and used the invariant form of Lyapunov exponent, where the proper
time is employed for the exponents as an invariant measure of time. Wu et al. \cite{xu1},
have tested the same problem using different approaches. More recent proposals and discussions
on that topic can be found in \cite{sukova,luke}. A review on Lyapunov exponent can be found
in \cite{kokos}.

The paper is organized as follows. In section 2, we provide the fundamentals of the
Lyapunov exponent.  In section 3, we shall completely describe the equatorial circular geodesics, both time-like 
and null cases for RN  BH and  also compute  the proper time Lyapunov exponent as well as coordinate time Lyapunov exponnent.
In section 4, we discuss  similar features for extremal RN space-time. In section 5, we relate the  QNMs of null 
circular geodesics  in terms of the Lyapunov exponent for a spherically symmetric RN space-time and Schwarzschild BH.  
In section 6, we present our conclusions.

\section{ Fundamentals of Lyapunov exponent:}

Lyapunov exponent in a classical phase space is a measure of the average rates of expansion and contraction 
of trajectories surrounding it. They are asymptotic quantities defined locally in state space, and describe the 
exponential rate at which a perturbation to a trajectory of a system grows or decays with time at a certain 
location in the state space. A positive Lyapunov exponent indicates a divergence between two nearby geodesics, the 
paths of such a system are extremely sensitive to changes of the initial conditions. A negative Lyapunov exponent 
implies a convergence between two nearby geodesics.

Let $x(t)$  i.e. $x(t=0)=x_{0}$ denote a trajectory of a system of equations governed by the following 
$n$-dimensional autonomous system \cite{naf}:
\begin{eqnarray}
\frac{dx}{dt} &=& F(x;M) ~.\label{dxdt}
\end{eqnarray}
The vector $x$ consists $n$ state variables, the function $F$ describes the non-linear evolution of the dynamical 
system and $M$ is a vector control parameter. Where $t$ is time parameter. . The solutions are fixed points or 
critical points when
$F(x;M) = 0$. Let its solution  for $M=M_{0}$ be $x_{0}$, where $x_{0}\in {\cal R}^{n}$ and $M_{0} \in {\cal R}^{m}$.  
To calculate the stability, we simply apply on $x(t)$, a small perturbation $y(t)$ and obtain
\begin{eqnarray}
x(t) &=& x_{0}+y(t) ~.\label{xt}
\end{eqnarray}
Substituting Eq. (\ref{xt}) into Eq. (\ref{dxdt}) yields
\begin{eqnarray}
\frac{dy}{dt} &=& F(x_{0}+y;M) ~.\label{dydt}
\end{eqnarray}
Note that the fixed point $x=x_{0}$ of Eq. (\ref{dxdt}) has been transformed into the fixed point $y=0$ of 
Eq.  (\ref{dydt}). Expanding Eq. (\ref{dydt}) in a Taylor series about $x_{0}$ and keeping only linear terms 
in the perturbation leads to
\begin{eqnarray}
\frac{dy}{dt} &=& F(x_{0};M_{0})+\frac{\partial F(x_{0};M_{0})}{\partial x} y
+O(\| y\|^{2}) ~.\label{tay}
\end{eqnarray}
or
\begin{eqnarray}
\frac{dy}{dt} &=& \frac{\partial F(x_{0};M_{0})}{\partial x} y = Ay ~.\label{tay1}
\end{eqnarray}
where the matrix $A$ is called Jacobian matrix. If the components of $F$ are
$F_{1}(x_{1},x_{2},x_{3}, ..., x_{n})$, $F_{2}(x_{1},x_{2},x_{3}, ..., x_{n})$,
 $F_{3}(x_{1},x_{2},x_{3}, ..., x_{n})$ then

\begin{eqnarray}
A=
\left(
  \begin{array}{cccc}
    \frac{\partial F_{1}}{\partial x_{1}} & \frac{\partial F_{1}}{\partial x_{2}} & ... &
    \frac{\partial F_{1}}{\partial x_{n}} \\
    \frac{\partial F_{2}}{\partial x_{1}} & \frac{\partial F_{2}}{\partial x_{2}} & ... &
    \frac{\partial F_{2}}{\partial x_{n}} \\
     .& . & . & . \\
     .& . & . & . \\
     .& . & . & . \\
    \frac{\partial F_{n}}{\partial x_{1}} & \frac{\partial F_{n}}{\partial x_{2}} &... &
    \frac{\partial F_{n}}{\partial x_{n}}\\
  \end{array}
\right) ~.\label{mata}
\end{eqnarray}

The eigen values of the constant matrix $A$ provide information about the local stability of the fixed 
point $x_{0}$. The  eigen values of $A$ are also known as characteristic exponents or Lyapunov exponents 
associated with $F$ at $(x_{0}, M_{0})$.

If we consider an initial deviation $y(0)$, its evolution is described by
\begin{eqnarray}
y(t)=\Phi(t) y(0) ~.\label{yt1}
\end{eqnarray}
where $\Phi(t)$ is the fundamental (transition) matrix solution of Eq. (\ref{tay1}) associated with the
trajectory say $x(t)$ which governs the dynamical equation  (\ref{dxdt}).
For an appropriate chosen the value of $y(0)$, the rate of exponential expansion or
contraction in the direction of $y(0)$ on the trajectory passing through
$x_{0}$ ( trajectory at $t=0$) is given by

\begin{eqnarray}
\lambda_{i} &=& \lim_{t \rightarrow \infty} \left( \frac{1}{t}\right)\ln
\left( \frac{\parallel y(t) \parallel}{ \parallel y(0) \parallel}\right) ~.\label{si}
\end{eqnarray}
where $\parallel \parallel$ denotes a vector norm. The asymptotic quantity $\lambda_{i}$ is
called the Lyapunov exponent.

If there exists a set of $n$ Lyapunov exponents associated with an n-dimensional autonomous
system and they can be ordered by size that is
\begin{eqnarray}
\lambda_{1}\geq \lambda_{2} \geq \lambda_{3}\geq, ... ,\geq \lambda_{n} ~.\label{le}
\end{eqnarray}
The set of n-numbers $\lambda_{i}$ is called the Lyapunov Spectrum.

Following Lyapunov \cite{lya}, the fundamental matrix $\Phi(t)$ is called regular if
\begin{eqnarray}
\lim_{t \rightarrow \infty} \ln \mid det \, \Phi(t)\mid ~.\label{det}
\end{eqnarray}
exist and is finite and if there exists a normal basis of the n-dimensional state
space such that
\begin{eqnarray}
\sum_{i=1}^{n}\lambda_{i}=\lim_{t \rightarrow \infty} \ln \mid det\, \Phi(t)\mid
~.\label{det1}
\end{eqnarray}
If $\Phi(t)$ is regular, then according to a theorem  by Oseldec\cite{osl} the asymptotic
quantity defined in Eq. (\ref{si}) exists and is finite for any initial deviation $y(0)$
belonging to the n-dimensional space.

The asymptotic quantity $\lambda_{i}$, given by Eq. (\ref{si}) is also known as a one-dimensional exponent.
For $p$-dimensions, the $p$-dimensional Lyapunov exponent $\lambda $ is defined as
\begin{eqnarray}
\lambda^{p} &=& \lim_{t \rightarrow \infty} \left( \frac{1}{t}\right)\ln
\left( \frac{\parallel y_{1}(t)\wedge y_{2}(t)\wedge ...\wedge y_{p}(t)\parallel}
{\parallel y_{1}(0)\wedge y_{2}(0)\wedge ... \wedge y_{p}(0)\parallel}\right) ~.\label{ly}
\end{eqnarray}
where $\wedge$ is an exterior or vector cross product.

In the next section, we will derive the expression for  Lyapunov exponent,  both using coordinate time and proper time.

\subsection{Lyapunov exponent and Radial Effective potential:}
Now we compute second derivative of the square of the radial component of the four velocity in terms
of the Lyapunov exponent. Therefore the Lagrangian of a test particle in the equatorial plane for
any static spherically symmetric space-time can be written as
\begin{eqnarray}
\cal L &=& \frac{1}{2}\left[g_{tt}\,{\dot{t}}^2+g_{rr}\,{\dot{r}}^2
+g_{\phi\phi}\,{\dot{\phi}}^2\right] ~.\label{lagg}
\end{eqnarray}
Now we define the canonical momenta as
\begin{eqnarray}
p_{q}&=& \frac{\partial {\cal L}}{\partial\dot{q}}~.\label{cm}
\end{eqnarray}
Using it, the generalized momenta can be derived as
\begin{eqnarray}
p_{t} &=& g_{tt}\,\dot{t}=-E =Const ~.\label{pt}\\
p_{\phi} &=& g_{\phi\phi}\,\dot{\phi}=L=Const ~.\label{pphi}\\
p_{r} &=& g_{rr}\, \dot{r}  ~.\label{pr}
\end{eqnarray}
Here $(\dot{t},~\dot{r},~\dot{\phi})$ denotes differentiation with respect to proper time($\tau$).
Again from the Euler-Lagrange equations of motion
\begin{eqnarray}
\frac{dp_{q}}{d\tau} &=& \frac{\partial{\cal L}}{\partial q}~.\label{el}
\end{eqnarray}
Using it, we get the non-linear differential equation in  two-dimensional phase space
with phase space variables $x_{i}(t)=(p_{r},~r)$.
\begin{eqnarray}
\frac{dp_{r}}{d\tau} &=& \frac{\partial {\cal L}}{\partial r} ~~\mbox{and}~~
\frac{dr}{d\tau} = \frac{p_{r}}{g_{rr}}~.\label{drdt}
\end{eqnarray}
Now linearizing the equation of motion about circular orbits of constant $r$, we get the infinitesimal
evolution matrix as
\begin{eqnarray}
M_{ij}=\left(
       \begin{array}{cc}
        0 & \frac{d}{dr}\left(\frac{\partial {\cal L}}{\partial r}\right)\\
        \frac{1}{g_{rr}} & 0 \\
      \end{array}
      \right)|_{r=r_{0}}  ~.\label{Mij}
\end{eqnarray}

For circular orbits of constant $r=r_{0}$ the characteristic values of the matrix gives the
information about the stability of the orbit. The eigen values of this matrix are called principal
Lyapunov exponent. Therefore the eigen values of the evolution matrix
along circular orbits can be written as
\begin{eqnarray}
\lambda^2= \frac{1}{g_{rr}}
\frac{d}{dr}\left(\frac{\partial {\cal L}}{\partial r}\right)|_{r=r_{0}}~.\label{eigen}
\end{eqnarray}
Again from Lagrange's equation of motion
\begin{eqnarray}
 \frac{d}{d\tau}\left(\frac{\partial {\cal L}}{\partial \dot{r}}\right)
 -\frac{\partial {\cal L}}{\partial r}=0~.\label{L}
\end{eqnarray}
Thus the Lyapunov exponent (which is the inverse of the instability time scale associated with the geodesic motions)
 in terms of the square of the radial velocity ($\dot{r}^2$) can be written as
\begin{eqnarray}
\frac{\partial {\cal L}}{\partial r}=\frac{1}{2g_{rr}}
\frac{d}{dr}\left(\dot{r}g_{rr}\right)^{2}~.\label{dLdr}
\end{eqnarray}
Finally,  the principal Lyapunov exponent can be rewritten as
\begin{eqnarray}
\lambda^2&=& \frac{1}{2} \frac{1}{g_{rr}}
\frac{d}{dr}\left[\frac{1}{g_{rr}}\frac{d}{dr}\left(\dot{r}g_{rr}\right)^{2}\right]
~.\label{pve}
\end{eqnarray}
Again for circular geodesics \cite{sch}
\begin{eqnarray}
\dot{r}^{2}=(\dot{r}^{2})'=0 ~.\label{cir}
\end{eqnarray}
where prime denotes the derivative with respect to $r$.
Thus,  for \emph{proper time}  Lyapunov exponent the equation (\ref{pve}) must be reduced to :
\begin{eqnarray}
\lambda_{p} & =& \pm \sqrt{\frac{(\dot{r}^{2})''}{2}}~.\label{pot}
\end{eqnarray}
 and  for coordinate time Lyapunov exponent \cite{car}   the equation (\ref{pve}) is given by
\begin{eqnarray}
\lambda_{c} & =& \pm \sqrt{\frac{(\dot{r}^{2})''}{2\dot{t}^{2}}}~~.\label{pot1}
\end{eqnarray}

The Lyapunov exponent must be in $\pm$ pairs to conserve the volume of the phase
space. From now on we shall take only positive Lyapunov exponent. The circular orbit is
unstable   when $\lambda_{p}$ or $\lambda_{c}$   is real, the circular orbit is stable
when the  $\lambda_{p}$ or $\lambda_{c}$  is imaginary and the circular orbit is marginally
stable when  $\lambda_{p}=0$ or $\lambda_{c}=0$ .

(Note that Cardoso et al.  \cite{car} have  derived a coordinate time
Lyapunov exponent, i.e. $\lambda_{c}  = \sqrt{\frac{V_{r}''}{2\dot{t}^2}}$ with $V_{r}=\dot{r}^{2}$ ).

The above expression for $\lambda$ is valid for any spherically symmetric
BH space-times \cite{pp1,remo,seta} i.e. (Schwarzschild, Reissner Nordstr{\o}m, Schwarzschild-de Sitter,
Schwarzschild-Anti-de Sitter, Reissner Nordstr{\o}m-de Sitter, Reissner Nordstr{\o}m-Anti de Sitter etc.).
Also it is valid for any axisymmetric space-time \cite{pp2,pp3,pp4}.

\subsection{Critical exponent and Radial Effective potential:}

Following Pretorius and Khurana \cite{pk}, we can define a critical exponent which is
the ratio of Lyapunov time scale or instability time scale $T_{\lambda}$ and Orbital
time scale $T_{\Omega}$ can be written as

\begin{eqnarray}
\gamma = \frac{\Omega}{2\pi\lambda}
=\frac{T_{\lambda}}{T_{\Omega}}=\frac{Lyapunov \, time scale}{Orbital \, time scale}~.\label{ce}
\end{eqnarray}
where  $T_{\lambda}=\frac{1}{\lambda}$ and $T_{\Omega}=\frac{2\pi}{\Omega}$. This is important for black-hole merger 
in the ring down radiation.
In terms of the 2nd derivative of the square of the radial velocity ($\dot{r}^2$), critical exponent
 can be written as
\begin{eqnarray}
\gamma_{p} &=&
\frac{1}{2\pi}\sqrt{\frac{2\Omega ^2}{(\dot{r}^{2})''}}~,\,\,
\gamma_{c} = \frac{1}{2\pi}\sqrt{\frac{2\dot{\phi}^2}{(\dot{r}^{2})''}}~
~.\label{cerd}
\end{eqnarray}
Here, $\Omega$ is the angular velocity.

We shall now calculate the equatorial circular geodesics for a spherically symmetric RN space-time.

\section{Equatorial Circular Geodesics in Spherically Symmetric RN Space-time:}

First, we shall consider a static, spherically symmetric, asymptotically flat solution of the coupled
Einstein-Maxwell equations in general relativity. They are described by the RN
space-time and the metric for such space-time is given by

\begin{eqnarray}
ds^2=-\left(1-\frac{2M}{r}+\frac{Q^{2}}{r^{2}}\right)dt^{2}+\frac{dr^{2}}{\left(1-\frac{2M}{r}+\frac{Q^{2}}{r^{2}}\right)}
+r^{2}\left(d\theta^{2}+\sin^{2}\theta d\phi^{2}\right) ~.\label{sph}
\end{eqnarray}

To compute the geodesics in the equatorial plane for this  space-time, we follow\cite{sch}. To determine
the geodesic motion of a test particle in this plane we set $\dot{\theta}=0$ and $\theta=constant=\frac{\pi}{2}$.

Therefore the necessary Lagrangian for this motion is given by
\begin{eqnarray}
\cal L &=& \frac{1}{2}\left[-\left(1-\frac{2M}{r}+\frac{Q^{2}}{r^{2}}\right)\,{\dot{t}}^2+\frac{{\dot{r}}^2}
{\left(1-\frac{2M}{r}+\frac{Q^{2}}{r^{2}}\right)}\,
+r^{2}\,{\dot{\phi}}^2\right] ~.\label{lags}
\end{eqnarray}

Using Eq. (\ref{cm}), the generalized momenta can be derived as
\begin{eqnarray}
p_{t} &=&-\left(1-\frac{2M}{r}+\frac{Q^{2}}{r^{2}}\right)\,\dot{t}=-E =Const ~.\label{pts}\\
p_{\phi} &=& r^{2}\,\dot{\phi}=L=Const ~.\label{pps}\\
p_{r} &=& \frac{\dot{r}}{\left(1-\frac{2M}{r}+\frac{Q^{2}}{r^{2}}\right)}\,   ~.\label{prs}
\end{eqnarray}
Here over dot denotes differentiation with respect to proper time($\tau$). Since the
Lagrangian does not depend on both `t' and `$\phi$', so $p_{t}$ and $p_{\phi}$ are
conserved quantities.
Solving Eq.  (\ref{pts}) and Eq. (\ref{pps}) for $\dot{t}$ and $\dot{\phi}$ we
find
\begin{eqnarray}
\dot{t}=\frac{E}{\left(1-\frac{2M}{r}+\frac{Q^{2}}{r^{2}}\right)} \,\, \mbox{and} \,\,
\dot{\phi}=\frac{L}{r^{2}} ~.\label{tdot}
\end{eqnarray}
where,  $E$ and $L$ are the energy and angular momentum per unit
rest mass of the test particle.

The normalization of the four velocity($u^{\mu}$) gives another integral equation
for the geodesic motion:
\begin{eqnarray}
g_{\mu\nu}u^{\mu}u^{\nu} &=& \epsilon ~.\label{norm}
\end{eqnarray}
or
\begin{eqnarray}
-E\dot{t}+L\dot{\phi}+\frac{{\dot{r}}^2}{\left(1-\frac{2M}{r}+\frac{Q^{2}}{r^{2}}\right)}&=& \epsilon ~.\label{norm1}
\end{eqnarray}
Here  $\epsilon=-1$ for time-like geodesics, $\epsilon=0$ for light-like geodesics and
$\epsilon=+1$ for space-like geodesics.
Substituting the value of $\dot{t}$ and $\dot{\phi}$ frm Eq.(\ref{tdot}) in Eq.(\ref{norm}),
we obtain the radial equation for spherically symmetric RN space-time:
\begin{eqnarray}
\dot{r}^{2}=E^{2}-\left(\frac{L^{2}}{r^{2}}-\epsilon \right)\left(1-\frac{2M}{r}+\frac{Q^{2}}{r^{2}}\right)~.\label{radial}
\end{eqnarray}

\subsection{Time-like Case:}
Now  the radial equation of the test particle for time-like circular geodesics \cite{pp1,remo} :
\begin{eqnarray}
\dot{r}^{2}=E^{2}- \left(1+\frac{L^{2}}{r^{2}}\right)\left(1-\frac{2M}{r}+\frac{Q^{2}}{r^{2}}\right)
~.\label{trn}
\end{eqnarray}

To investigate the circular geodesic motion of the test particle in the Einstein -Maxwell gravitational field , 
for circular geodesics we must have of constant $r=r_{0}$  and using the condition for circular orbit(\ref{cir}), we 
get the energy and angular momentum per unit mass of the test particle for time-like orbit are
\begin{eqnarray}
E_{0}^{2} &=&  \frac{\left(r_{0}^{2}-2Mr_{0}+Q^{2}\right)^{2}}{r_{0}^{2}(r_{0}^{2}-3Mr_{0}+2Q^{2})}~~\mbox{and}~~
L_{0}^{2}= \frac{r_{0}^{2}\left(Mr_{0}-Q^{2}\right)}{r_{0}^{2}-3Mr_{0}+2Q^{2}}~.\label{angn}
\end{eqnarray}

Circular motion is possible when both energy and angular momentum are real and finite, therefore we
must have $r_{0}^{2}-3Mr_{0}+2Q^{2}>0$ and $r_{0}>\frac{Q^{2}}{M}$, and the angular frequency  at
$r=r_{0}$ is
\begin{eqnarray}
\Omega_{0}=\frac{\dot{\phi}}{\dot{t}}=\frac{\sqrt{\left(Mr_{0}-Q^{2}\right)}}{r_{0}^{2}} ~.\label{omt2}
\end{eqnarray}

Hence for non-extremal RN BH, the proper time  Lyapunov exponent and coordinate time Lyapunov exponent
are
\begin{eqnarray}
{\lambda_{p}}^{RN}&=& \sqrt{\frac{-(Mr_{0}^{3}-6M^{2}r_{0}^{2}+9MQ^{2}r_{0}-4Q^{4})}{r_{0}^{4}
(r_{0}^{2}-3Mr_{0}+2Q^{2})}}~.\label{marn}\\
{\lambda_{c}}^{RN}&=&  \sqrt{\frac{-(Mr_{0}^{3}-6M^{2}r_{0}^{2}+9MQ^{2}r_{0}-4Q^{4})}{r_{0}^{6}
}} ~.\label{marc}
\end{eqnarray}

Thus the time-like circular geodesics of non-extremal RN BH is
stable when $Mr_{0}^{3}-6M^{2}r_{0}^{2}+9MQ^{2}r_{0}-4Q^{4}>0$ such that  $\lambda_{p}$ or $\lambda_{c}$
is imaginary, the circular geodesics is unstable when $Mr_{0}^{3}-6M^{2}r_{0}^{2}+9MQ^{2}r_{0}-4Q^{4}<0$ i.e.
$\lambda_{p}$ or $\lambda_{c}$  is real and the time-like circular geodesics is marginally stable
when $Mr_{0}^{3}-6M^{2}r_{0}^{2}+9MQ^{2}r_{0}-4Q^{4}=0$ such that  $\lambda_{p}$ or $\lambda_{c}$
is equal to zero.

For completeness, we have also  computed the ratio of $\frac{\lambda_{p}}{\lambda_{c}}$ for
RN spacetime which is given by
\begin{eqnarray}
\frac{\lambda_{p}}{\lambda_{c}} &=& \frac{r_{0}}{\sqrt{r_{0}^2-3Mr_{0}+2Q^2}}
\end{eqnarray}
For extremal RN BH,  this ratio is:
\begin{eqnarray}
\frac{\lambda_{p}}{\lambda_{c}} &=& \frac{r_{0}}{\sqrt{(r_{0}-M)(r_{0}-2M)}}
\end{eqnarray}
For Schwarzschild BH, this ratio has been reduced to
\begin{eqnarray}
\frac{\lambda_{p}}{\lambda_{c}} &=& \frac{\sqrt{r_{0}}}{\sqrt{r_{0}-3M}}
\end{eqnarray}
Now we will see the variation of  the ratio of $\lambda_{p}$ and $\lambda_{c}$ in graphically (See Fig. 1) for
Schwarzschild BH $(Q=0)$ and RN BH .

\begin{center}
\scalebox{0.5}{\includegraphics{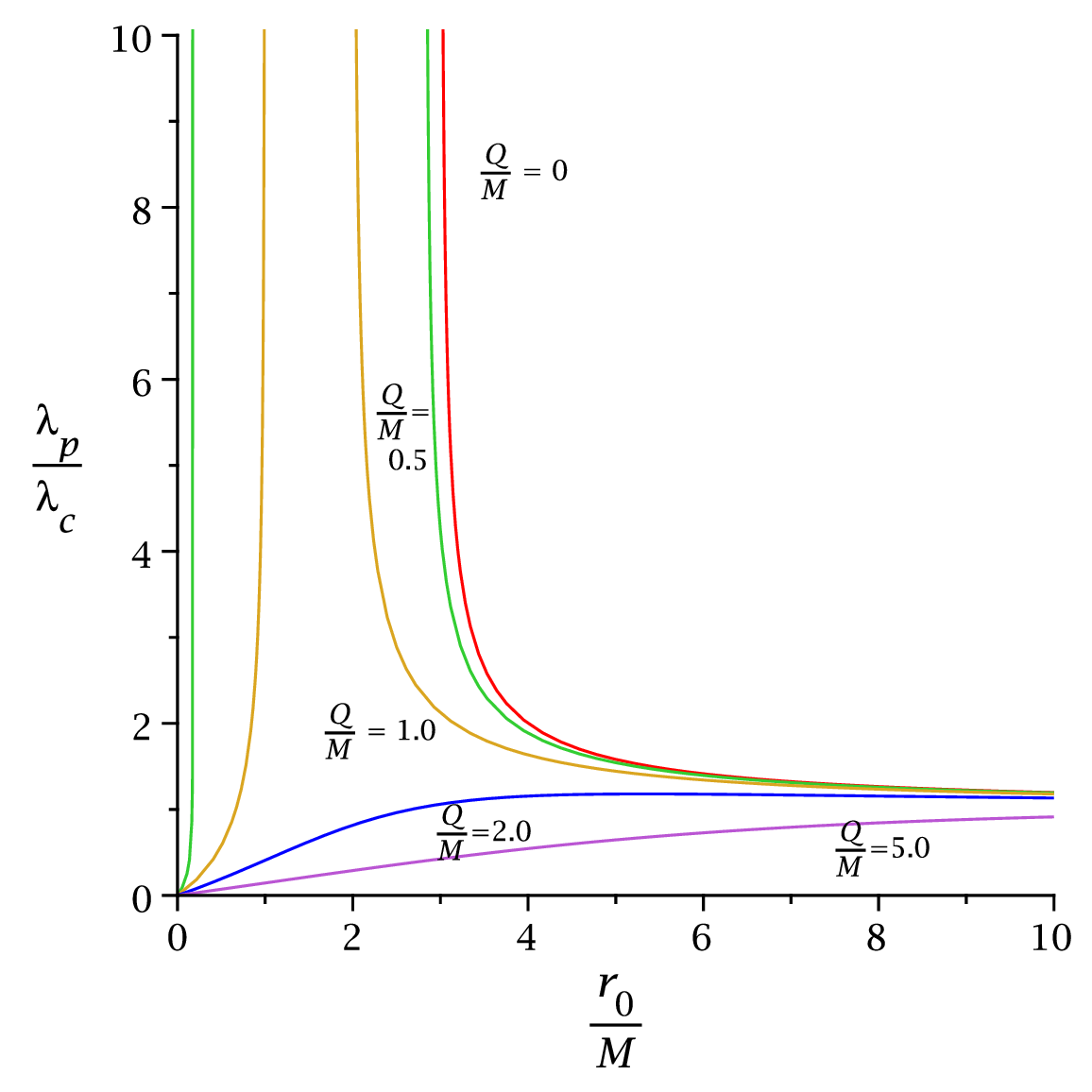}}\\
\vspace{0.02cm}
\end{center}
Fig. 1. The  variation  of  $\frac{\lambda_{p}}{\lambda_{c}}$
with $\frac{r_{0}}{M}$ for RN BH.
The range of validity for the  charge to mass ratio $\frac{Q}{M}$ is $0 \leq \frac{Q}{M} \leq 1$.
In the graph, red color indicates for  the value of charge   $Q= 0$ , green color indicates for the value of
charge $Q= 0.5M$, yellow color  indicates for the value of charge  $Q= 1.0M$, blue color indicates  for the
 value of charge $Q= 2.0M$ and violet color indicates for  $Q= 5.0M$. \\

It can be easily seen from Fig. 1 that,  the  ratio of $\lambda_{p}$ and $\lambda_{c}$ varies from orbit to orbit for
different charge parameters.

Thus the  solution of the system $\dot{r}^{2}=(\dot{r}^{2})'=(\dot{r}^{2})''=0$
gives the radius of ISCO at $r_{0}=r_{ISCO}$ for non-extremal RN BH
which is given by
\begin{eqnarray}
\frac{r_{ISCO}}{M}&=& 2+Z^{\frac{1}{3}}+\frac{4-3\left(\frac{Q}{M}\right)^2}{Z^{\frac{1}{3}}} ~.\label{isco}
\end{eqnarray}
where
\begin{eqnarray}
Z &=& 8-9\left(\frac{Q^{2}}{M^{2}}\right)+2\left(\frac{Q^{4}}{M^{4}}\right)+
\sqrt{5\left(\frac{Q^{4}}{M^{4}}\right)-9\left(\frac{Q^{6}}{M^{6}}\right)+4\left(\frac{Q^{8}}{M^{8}}\right)}
\end{eqnarray}
Since Schwarzschild space-time is the special case of RN space-time which occurs
in the limit $Q=0$ and  therefore we find the radius of ISCO,  $r_{ISCO}=6M$.

Now the reciprocal of critical exponent for RN BH   is given by
\begin{eqnarray}
\frac{1}{\gamma_{p}}&=&\frac{T_{\Omega}}{T_{\lambda}}=
2\pi \sqrt{\frac{-(Mr_{0}^{3}-6M^{2}r_{0}^{2}+9MQ^{2}r_{0}-4Q^{4})}{(Mr_{0}-Q^2)
(r_{0}^{2}-3Mr_{0}+2Q^{2})}}~.\label{cern}
\end{eqnarray}
For any unstable circular orbit, $T_{\Omega}>T_{\lambda}$ i.e. Lyapunov time scale
is shorter than the gravitational time scale making the instability observationally
relevant\cite{cl}.

\emph{Special case:}

For Schwarzschild BH $Q=0$, the Lyapunov exponent is
\begin{eqnarray}
{\lambda_{p}}^{Sch} &=& \sqrt{-\frac{M(r_{0}-6M)}{r_{0}^{3}(r_{0}-3M)}}~.\label{marg3}\\
{\lambda_{c}}^{Sch} &=& \sqrt{-\frac{M(r_{0}-6M)}{r_{0}^{4}}}~.\label{marg3a}
\end{eqnarray}

The reciprocal of critical exponent for Schwarzschild BH is given by
\begin{eqnarray}
\frac{1}{\gamma_{p}} &=& \frac{T_{\Omega}}{T_{\lambda}}=
2\pi \sqrt{\frac{-(r_{0}-6M)}{(r_{0}-3M)}}~.\label{cesc}
\end{eqnarray}

For any unstable circular orbit say for $r_{0}=4M$, $\gamma$ becomes $\frac{1}{2\sqrt{2}\pi}$.
Therefore $T_{\lambda}<T_{\Omega}$, i.e. Lyapunov time scale is shorter than the gravitational
time scale \cite{cl}.

\subsection{Null case:}
As there is no proper time for photons, we have to only compute the coordinate
time Lyapunov exponent. To proceed this for null geodesics, the radial equation is given by

\begin{eqnarray}
\dot{r}^{2}=E^{2}- \frac{L^{2}}{r^{2}}\left(1-\frac{2M}{r}+\frac{Q^{2}}{r^{2}}\right)~.\label{nc}
\end{eqnarray}

Therefore the energy and angular momentum evaluated at $r=r_{c}$ for circular null
geodesics are
\begin{eqnarray}
\frac{E_{c}}{L_{c}}=\pm \sqrt{\frac{r_{c}^{2}-2M r_{c}+Q^{2}}{r_{c}^{4}}} ~~\mbox{and}~~
r_{c}^{2}-3M r_{c}+2Q^{2}= 0 ~.\label{n1}
\end{eqnarray}

After introducing the impact parameter $D_{c}=\frac{L_{c}}{E_{c}}$, the above equation
is reduced to
\begin{eqnarray}
\frac{1}{D_{c}}=\frac{E_{c}}{L_{c}}=\sqrt{\frac{Mr_{c}-Q^{2}}{r_{c}^{4}}}=\Omega_{c}=
\frac{\dot{\phi}}{\dot{t}} ~.\label{rdc1}
\end{eqnarray}
where $\Omega_{c}$ is the angular frequency measured by an asymptotic observers at infinity.

Solving equation (\ref{n1}) we obtain the radius of null circular orbits \cite{cve} as
\begin{eqnarray}
{r_{c}}_{\pm} &=& \frac{3M}{2}\left[1\pm\sqrt{1-\frac{8}{9}(\frac{Q^{2}}{M^{2}})}\right] ~.\label{rnt}
\end{eqnarray}

Using (\ref{pot}), the Lyapunov exponent for null circular geodesics is given by
\begin{eqnarray}
{\lambda_{c}}^{null}\mid_{RN} &=& \sqrt{\frac{(Mr_{c}-Q^{2})(3Mr_{c}-4Q^{2})}{r_{c}^{6}}}
~.\label{tn5}
\end{eqnarray}

So the circular geodesics $r_{c}={r_{c}}_{+}$ and $r_{c}={r_{c}}_{-}$ are unstable
since ${\lambda_{c}}^{null}\mid_{RN}$ is real.

For Schwarzschild BH $Q=0$, the Lyapunov exponent reads
\begin{eqnarray}
{\lambda_{c}}^{null} \mid_{Sch} &=&  \frac{\sqrt{3}M}{r_{c}^{2}} ~.\label{tn6}
\end{eqnarray}
It can be easily check that for $r_{c}=3M$, ${\lambda_{c}}^{null}\mid_{Sch}$ is real, that means
Schwarzschild photon sphere is unstable.

\section{Extremal RN Space-time:}

\subsection{Lyapunov exponent and Equation of ISCO:}

For extremal RN BH, the radial equation of the test particle for time-like circular
geodesics is
\begin{eqnarray}
\dot{r}^{2}=E^{2}- \left(1+\frac{L^{2}}{r^{2}}\right)\left(1-\frac{M}{r}\right)^{2}
~.\label{trc}
\end{eqnarray}

Thus the Lyapunov exponent  becomes
\begin{eqnarray}
{\lambda_{p}}^{ex} &=& \frac{\sqrt{M(r_{0}-M)}}{r_{0}^2\sqrt{(r_{0}-2M)}}\sqrt{-\left(r_{0}-4M\right)}
~.\label{marg2}\\
{\lambda_{c}}^{ex}&=& (r_{0}-M)\sqrt{\frac{-M(r_{0}-4M)}{r_{0}^{6}
}} ~.\label{marx}
\end{eqnarray}

So,  the time like circular geodesics of extremal RN BH is
stable when $r_{0}>4M$ such that  $\lambda_{p}$ or $\lambda_{c}$  is imaginary, the circular geodesics is
unstable when $2M<r_{0}<4M$  i.e  $\lambda_{p}$ or $\lambda_{c}$  is real and the time-like circular
geodesics is marginally stable when $r_{0}=4M$ such that  $\lambda_{p}$ or $\lambda_{c}$  is  equal to zero.

\subsection{Lyapunov exponent and Null Circular Geodesics:}

Analogously, using (\ref{pot}) the Lyapunov exponent for null circular geodesics:
\begin{eqnarray}
{\lambda_{c}}^{exn} &=&  \sqrt{\frac{M^{2}(r_{c}-M)(3r_{c}-4M)}{r_{c}^{6}}}
~~.\label{tn7}
\end{eqnarray}
So the circular geodesics $r_{c}=2M$ are unstable since $\lambda_{c}^{exn}$ is real.
\emph{Note that},  for extremal BH, this result is valid for only single null
geodesics i.e. $r_{0}\neq r_{c}$. For $r_{0}=r_{c}=M$, the Lyapunov exponent becomes
zero i.e. $\lambda_{p}^{ex}=\lambda_{c}^{ex}=\lambda_{c}^{exn}=0$.

Now we shall make a link between Lyapunov exponent of  null circular geodesics
and QNM for RN BH in the eikonal limit.

\section{Null Circular Geodesics and QNM for RN BH in the Eikonal limit:}

This section is devoted to study the QNM frequencies for RN BH in the eikonal limit
following the work by\cite{car}. It is well known that the unstable null circular
geodesics is very useful to determine the characteristic modes of BH, which is so
called the QNMs\cite{kostas,nollert,konoplya}. To
compute it, we first consider the wave equation for a massless scalar field in the back ground
of RN space-time may be cast in the form :

\begin{eqnarray}
\frac{d^2X}{d{r_{\ast}}^2}+Q_{0}X &=&0~ \label{qn1}
\end{eqnarray}
where,
\begin{eqnarray}
Q_{0} &=& \omega^2-V_{s}(r) \,\, \mbox{and}   \,\,
V_{s}(r)= \frac{\ell(\ell+1) }{r^2}+\frac{2(Mr-Q^2)(r^2-2Mr+Q^2)}{r^{6}}  ~ \label{qqn2}
\end{eqnarray}
Here, $\ell$ denotes the spherical harmonic index and $r_{\ast}$ is the ``tortoise''
coordinates, ranging from $-\infty$ to  $+\infty$.

In the eikonal limit i.e. $\ell\rightarrow \infty$, we get

\begin{eqnarray}
Q_{0} &\approx & \omega^2-\frac{\ell}{r^2} \left(1-\frac{2M}{r}+\frac{Q^{2}}{r^{2}}\right)  ~ \label{eik}
\end{eqnarray}
Using Eq. (\ref{eik}), one may able to find the maximum value of $Q_{0}$ which
occurs at $r=r_{z}$:
\begin{eqnarray}
r_{z}^2-3Mr_{z}+2Q^2 &=& 0 ~.\label{nq1}
\end{eqnarray}
Again we know, for null circular geodesics the radial equation is
determined by the Eq.(\ref{n1}). Therefore unstable circular orbits could be determined by the
condition $\dot{r}^{2}=0$, leading to
\begin{eqnarray}
r_{c}^2-3Mr_{c}+2Q^2 &=& 0 ~.\label{nnq2}
\end{eqnarray}
Since the  maximum value of $Q_{0}$ and the location of the null circular geodesics are
coincident at $r_{z}=r_{c}$. Therefore from Eq.(\ref{qn1}), one may find the QNM
conditions\cite{schutz,will,iyer}:
\begin{eqnarray}
\frac{Q_{0}(r_{z})}{\sqrt{-2\frac{d^2Q_{0}}{d{r_{\ast}}^2}}}= i\left(n+1/2\right) ~.\label{nq2}
\end{eqnarray}
Eq.(\ref{nq1}) is evaluated at the extremum of $Q_{0}$ i.e., the point $r_{0}$ at which
$\frac{dQ_{0}}{d{r_{\ast}}}=0$. Therefore in the large-$\ell$ limit Eq.(\ref{nq2}) gives
\begin{eqnarray}
\omega_{QNM}=\ell \sqrt{\frac{Mr_{c}-Q^{2}}{r_{c}^{4}}} - i\left(n+1/2\right) \sqrt{\frac{(Mr_{c}-Q^{2})(3Mr_{c}-4Q^{2})}{r_{c}^{6}}} ~.\label{nq3}
\end{eqnarray}

The significance of the above Eq.(\ref{nq3})  is that in the eikonal limit, the real and imaginary
parts of the QNMs of RN BH are given by the frequency and instability time scale of the
unstable null circular geodesics. This is one of the key results of the paper.

It should be noted that for extremal RN BH, since ${\lambda_{c}}^{null}=0$ for  $r_{0}=r_{c}=M$, therefore
the value of $\omega_{QNM}$ becomes:.
\begin{eqnarray}
\omega_{QNM}=0 ~.\label{nq4}
\end{eqnarray}
 For Schwarzschild BH,  in the eikonal limit  the frequency of QNM is given by
\begin{eqnarray}
\omega_{QNM} &=& \ell \sqrt{\frac{M}{r_{c}^{3}}} - i\left(n+1/2\right) \frac{\sqrt{3}M}{r_{c}^{2}}
\end{eqnarray}
Thus by calculating the Lyapunov exponent, which is the reciprocal of the instability time scale associated
with the geodesic motion, we  found that, in the eikonal limit, the  frequency of QNMs of Schwarzschild BH
could be determined by the parameters of the null circular geodesics.

\section{Discussion}
In this article, we have used the Lyapunov exponent to give a full descriptions of  time-like circular  geodesics and
null circular geodesics in a spherically symmetric  RN BH  space-time.  Then we explicitly derived the proper time Lyapunov
exponent and coordinate time Lyapunov exponent for RN BH.
We  found that the ratio of proper time LE and coordinate time LE for RN space-time is
$\frac{\lambda_{p}}{\lambda_{c}}=\frac{r_{0}}{\sqrt{r_{0}^2-3Mr_{0}+2Q^2}}$ for time
like circular geodesics and for Schwarzschild BH the ratio is being
$\frac{\lambda_{p}}{\lambda_{c}}=\frac{\sqrt{r_{0}}}{\sqrt{r_{0}-3M}}$. This ratio also
varies from orbit to orbit and may not contain any generic information.
For example, for Schwarzschild BH at ISCO the ratio is
$\frac{\lambda_{p}}{\lambda_{c}}\mid_{r_{ISCO}=6M}=\sqrt{2}$ and at
marginally bound circular orbit (MBCO) the ratio is calculated
to be $\frac{\lambda_{p}}{\lambda_{c}}\mid_{r_{mb}=4M}=2$.
Similarly, for  extremal RN BH the ratio is at ISCO
$\frac{\lambda_{p}}{\lambda_{c}}\mid_{r_{ISCO}=4M}=\frac{2\sqrt{2}}{\sqrt{3}}$
and at MBCO is $\frac{\lambda_{p}}{\lambda_{c}}\mid_{r_{mb}=\frac{3+\sqrt{5}}{2}M}=\frac{3+\sqrt{5}}{2}$.

We  further showed that the Lyapunov exponent can be used to determine the stability and instability of
equatorial circular geodesics, both time-like and null case for RN BH space-time.
Finally, we computed  the QNM frequencies for RN BH in the eikonal limit.  We  found  that  in the eikonal limit,
the real and imaginary parts of the QNMs of RN BH is given by the frequency and instability time scale of the
unstable null circular geodesics.  For Schwarzschild BH, in the eikonal limit,  the real part of the complex QNM frequencies
 are detemined by the angular velocity of the null circular geodesics and imaginary part is  related to the coordinate time
Lyapunov exponent of null circular geodesics. 

Besides the theory of Lyapunov exponent has important applicatin in the study of critical phenomena in BH binaries 
\cite{berti,sper,mcw}. 
It also plays a crucial role for the physical understanding of ring-down radiation, interpretation of numerical simulations of 
BH merger and gravitational wave data analysis. It also plays a crucial role.

\section*{Acknowledgements}
The author would like to thank Prof. Parthasarathi Majumdar of R.M.V.U for helpful discussions. He
also wishes to thank Prof. Kumar Shwetketu Virbhadra for his comments and
suggestions regarding the ``Photon Sphere''. Finally, he would like to thank the anonymous 
referee for his valuable suggestions.

\bibliography{apssamp}% Produces the bibliography via BibTeX.

\begin{thebibliography}{99}
\bibitem{lya} A. M. Lyapunov, \textit{ The General Problem of the Stability of Motion}, Taylor and Francis, London (1992).

\bibitem{car} V. Cardoso et al. , \textit { Phys. Rev.} {\bf D 79} (2009) 064016.

\bibitem{berti} E.~Berti,  arXiv:1410.4481v2.

\bibitem{osl} V.~I.~Oseldec, \textit{ Trans. Moscow. Math. Soc.} {\bf 19}, 197 (1968).

\bibitem{clv} N.~J.~Cornish, \textit{ Phys. Rev.} {\bf D 64}, 084011 (2001).

\bibitem{cl} N. ~J. ~Cornish and J.~J.~Levin, \textit{ Class. Quant. Grav. } {\bf 20}, 1649(2003).

\bibitem{karas} V. Karas and D. Vokrouhlicky, \textit{ Gen.  Relativ.  Gravit. } {\bf 24}, 729 (1992).

\bibitem{mott} A.~E.~Motter, \textit{ Phys. Rev. Lett.} {\bf  91}, 231101 (2003).

\bibitem{xu} X. Wu, T. Y. Huang, \textit{ Physics Letters} {\bf A 313},  77 (2003).

\bibitem{sota} Y. ~Sota, S. Suzuki and K.I. Maeda, \textit{ Class. Quant. Grav.} {\bf 13}, 1241 (1996).

\bibitem{xu1} X. Wu, T. Y. Huang and H. Zhang \textit{ Phys. Rev. } {\bf D 74}, 083001 (2006).

\bibitem{sukova}   Sukov\'{a} and O. Semer\'{a}k \textit{Mon. Not. R. Astro. Soc.} {\bf 436}, 978 (2013).

\bibitem{luke} G.~Lukes ~Gerakopoulos, \textit{ Phys. Rev. } {\bf D 89}, 043002  (2014).

\bibitem{kokos} C. Skokos, \textit{ Lect. Notes Phys.} {\bf 790} , 63  (2010).

\bibitem{naf} A. H. Nayfeh and B. Balachandran, \textit{ Applied Nonlinear Dynamics }, Wiley-Vch Verlag GmbH  Co., (2004).

%\bibitem{ott} E. Ott, \textit{ Chaos in Dynamical Systems}, Cambridge University Press, Cambridge (1994).

\bibitem{pp1} P. Pradhan, P. Majumdar, \textit{ Physics Letters } {\bf A 375} (2011) 474-479.

\bibitem{remo} D. Pugliese, H. Quevedo, R. Ruffini, \textit{ Phys. Rev.} {\bf D 83} (2011) 024021.

\bibitem{seta} M. R. ~Setare and D. ~Momeni, \textit{ Int~ J~ Theor~ Phys.} {\bf 50}, 106-113 (2011).

\bibitem{pk} F.~Pretorius and D.~Khurana, \textit{ Class. Quant. Grav.}  {\bf 24}, S83 (2007).

\bibitem{sch} S. Chandrashekar, \textit{ The Mathematical Theory of Black Holes}, Clarendon Press, Oxford (1983).

\bibitem{cve} C. M.~Claudel, K. S.~Virbhadra and G. F. R.~Ellis,~ \textit{ Journal of~ Math~ Phys.} {\bf 42}, 818 (2001).

\bibitem{pp2} P. Pradhan, P. Majumdar, \textit{ Eur. Phys. J. C} {\bf 73: 2470} (2013).

\bibitem{pp3} P. Pradhan, \textit{ ISCO, Lyapunov exponent and Kerr-Newman space-time;}, arXiv:1212.5758 [gr-qc].

\bibitem{pp4} P. Pradhan, \textit{ Eur. Phys. J. C} {\bf 73: 2477} (2013).

\bibitem{kostas} K. D. Kokkotas and  B. Schmidt, \textit{Living Rev. Relativity} {\bf 2,} {2} (1999).

\bibitem{nollert} H. P. ~Nollert, \textit{ Class. Quant. Grav.}  {\bf 16}, R159  (1999).

\bibitem{konoplya} R. A. Konoplya, \textit{Rev. Mod. Phys. } {\bf 83}, 793 (2011).

\bibitem{schutz} B. F. Schutz and C. M. Will, \textit{Astrophys. J.} {\bf 291} {L33} (1985).

\bibitem{will} S. Iyer and  C. M. Will,  \textit{ Phys. Rev. } {\bf D 35}, 3621 (1987).

\bibitem{iyer} S. Iyer,  \textit{ Phys. Rev. } {\bf D 35}, 3632 (1987).

%\bibitem{berti} E.~Berti,  arXiv:1410.4481v2.

\bibitem{sper} E. Berti et al. , \textit{ Phys. Rev.} {\bf D76}, 064034 (2007).

\bibitem {mcw} J. G. Baker et al. , \textit{ Phys. Rev.} {\bf D78}, 044046 (2008).

\end{thebibliography}

\end{document}